\documentclass{article}
\usepackage[T1]{fontenc}
\usepackage[utf8]{inputenc}
\usepackage{ismir}
\makeatletter
\renewcommand{\permission}{}
\makeatother
\usepackage{amsmath,amssymb,cite,url,graphicx,booktabs,array,placeins}
\usepackage[bookmarks=false,hidelinks]{hyperref}
\newcommand{\faisonic}{FaiSonic~MMD}
\title{\faisonic: A Popular Music Dataset with Temporally Aligned Multidimensional Annotations for Multitask Music Analysis}
\multauthor
  {Tieyao Zhang$^{1,4}$ \hspace{0.6cm} Yuke Liu$^{1,4}$ \hspace{0.6cm} Jiaxing Yu$^2$}
  {\textbf{Xinda Wu$^2$ \hspace{0.6cm} Genfang Chen$^{1,3}$}\\[0.3em]
  {\normalsize $^1$ Zhejiang Conservatory of Music, Hangzhou, China}\\
  {\normalsize $^2$ Zhejiang University, Hangzhou, China}\\
  {\normalsize $^3$ ZJU-ZJCM MusicAI Lab, Hangzhou, China}\\
  {\normalsize $^4$ Hangzhou FaiSonic Intelligent Music Technology Co., Ltd., Hangzhou, China}\\
  {\ttfamily\small tieyaozhang@stu.zjcm.edu.cn, keke@stu.zjcm.edu.cn, yujx@zju.edu.cn,}\\
  {\ttfamily\small wuxinda@zju.edu.cn, cgf@zjcm.edu.cn}}
\def\authorname{T. Zhang, Y. Liu, J. Yu, X. Wu, and G. Chen}
\hypersetup{pdfauthor={Tieyao Zhang, Yuke Liu, Jiaxing Yu, Xinda Wu, Genfang Chen},pdftitle={FaiSonic MMD: A Popular Music Dataset with Temporally Aligned Multidimensional Annotations for Multitask Music Analysis}}
\begin{document}
\maketitle
\thispagestyle{empty}

\begin{abstract}
Temporal analysis in music information retrieval concerns the positions of musical events and structures within recordings. Joint modeling has demonstrated the value of combining musical descriptions, but existing resources differ in joint coverage and expert verification across dimensions, constraining detailed joint supervision and the analysis of relationships on the same recordings. To support multitask modeling and the study of relationships among musical dimensions in popular music, we introduce FaiSonic Multidimensional Music Dataset (FaiSonic~MMD), containing 500 songs totaling approximately 32.86 hours, comprising 250 Chinese-language and 250 English-language songs. Professional musicians annotate phrases, sections, chords, and rhythm in real recordings, followed by independent review. Events and intervals across dimensions share a recording-based time reference while retaining their respective musical granularity and boundaries, allowing direct comparison across musical descriptions. We describe the annotations, collection procedure, and statistics, and establish a comprehensive multitask baseline for phrase boundaries, section boundaries, section functions, chords, beats, and downbeats. We report reference performance for all prediction targets. FaiSonic~MMD provides shared expert annotations and accompanying benchmarks for temporal music analysis.

\end{abstract}

\section{Introduction}
Temporal analysis in music information retrieval focuses on the positions and organization of musical events and structures within recordings. Beat tracking, chord recognition, phrase segmentation, and section analysis describe rhythm, harmony, and structure at different scales~\cite{ref1,ref6,ref31}. These tasks support song navigation, music transcription, and content analysis~\cite{ref1,ref5}. Research typically defines training targets and evaluation protocols for individual tasks, while relationships such as those between chord changes and metrical positions, or phrase organization and section function, offer opportunities for analysis across dimensions~\cite{ref15,ref29}. Joint modeling and relationship analysis may inform the use of complementary information, the interpretation of prediction errors, and the development of individual tasks. Extending such research to more dimensions requires attention to task coverage, supervision sources, and the organization of temporal annotations.

Previous work has improved the performance of some temporal tasks through joint training, providing evidence for using relationships among musical dimensions~\cite{ref15}. Representative methods nevertheless address particular task combinations: All-In-One jointly predicts beats, downbeats, structural boundaries, and functional labels~\cite{ref15}; Beat-U adds chords, but its structural target is limited to boundaries and its multidimensional training material comes from an internal J-Pop collection~\cite{ref31}. Neither includes independent phrases as a prediction target. Differences in task combinations and the associated data leave scope for broader research on relationships among dimensions.

SheetSage2 extends unified transcription to rhythm, harmony, melody, and form, identifying the scarcity of complete, accurately aligned audio--score pairs as a training bottleneck~\cite{ref29}. It expands supervision through automatically annotated MIDI and rendered audio, combining real-recording annotations, structured decoding, and teacher-generated pseudo-labels to handle incomplete or noisy supervision~\cite{ref29}. This approach shows how automated supervision can support broader joint analysis, while highlighting a complementary data need: multidimensional temporal references in real recordings, annotated and reviewed song by song by experts, for detailed supervision and the study of musical relationships.

Further joint research therefore requires annotations that cover the relevant dimensions together and verify both musical content and temporal positions~\cite{ref5,ref29}. Separate datasets for individual tasks do not necessarily provide directly comparable multidimensional information for the same recordings; existing resources also emphasize different combinations of dimensions~\cite{ref4,ref5,ref30}. Temporal annotations covering the same songs allow tasks to share training material and events and intervals to be compared directly across dimensions. Preserving each dimension's own granularity and boundaries further enables the study of relationships among local phrases, section functions, harmonic changes, and metrical positions.

To address these needs, we introduce FaiSonic Multidimensional Music Dataset (FaiSonic~MMD), providing expert annotations of phrases, sections, chords, and rhythm for Chinese- and English-language popular recordings. This combination covers metrical positions, harmonic content, and musical organization at local and larger scales. Initial annotation and independent review preserve each dimension's boundaries and granularity on a shared recording timeline, providing a basis for joint modeling and the analysis of relationships among dimensions. The data and accompanying code are publicly available through the project repository.\footnote{\url{https://github.com/Kreuzter0421/FaiSonic-MMD}}

To support the joint use of all four annotation dimensions, we establish a comprehensive multitask baseline with a complete training, inference, and evaluation pipeline. Our main contributions are:
\begin{enumerate}
\item \textbf{An expert-annotated dataset covering multiple dimensions of temporal music analysis.} FaiSonic~MMD contains 500 popular songs totaling approximately 32.86 hours, comprising 250 Chinese-language and 250 English-language songs. All songs have phrase, section, chord, and rhythm annotations produced by experts and independently reviewed.
\item \textbf{Temporal alignment across musical dimensions.} Using the same recording as a time reference, we organize phrase, section, chord, and rhythm events and intervals on a shared timeline while preserving each dimension's musical granularity and independent boundaries. This supports direct comparison of musical information at different scales and provides a basis for studying relationships among dimensions.
\item \textbf{A comprehensive multitask baseline covering all annotation dimensions.} We provide a complete implementation from shared audio representations and task-specific branches to structured decoding, and report reference results for phrase boundaries, section boundaries and functions, chords, beats, and downbeats to support the joint use of the dataset.
\end{enumerate}

\section{Related Work}
\subsection{Music Annotation Resources}
Temporal annotation resources describe musical content through events, intervals, and labels, with analytical goals determining the descriptive perspective and temporal scale. Peeters and Deruty distinguish structural descriptions based on similarity, function, instrumentation, and other perspectives~\cite{ref2}; hierarchical structure evaluation further considers segmentation and containment across scales~\cite{ref6}. SALAMI organizes structural similarity and function at multiple scales. Its original paper reports 1,383 recordings and more than 2,400 structural annotations spanning multiple genres~\cite{ref3}. Its finer structural level often corresponds to short phrases, but may also include motives and other units~\cite{ref1,ref6}. Multiple annotators may describe the same recording, enabling the study of alternative musical segmentations.

SongFormer provides the large-scale training collection SongFormDB and the evaluation set SongFormBench~\cite{ref7}. SongFormDB contains over 14,000 songs, integrating human structural annotations, partial excerpt annotations, automatically aligned data, and model-generated labels; some songs have annotations only for excerpts. SongFormBench consists of 200 songs from Harmonix and 100 Chinese-language songs, cross-checked by experts using audio, existing labels, and other supporting information for structure evaluation.

Harmonic resources can also provide descriptions linked to musical structure. Version 2.0 of McGill Billboard contains 740 distinct songs and 890 chart-sampled entries, organizing chords, phrases, and larger-scale sections within the same annotations~\cite{ref4}. Experts transcribe chord content in terms of bars and beat positions. In the accompanying LAB files, only the first chord of each phrase is manually aligned; other chord times are linearly interpolated under a constant-tempo assumption within the phrase. RWC-Pop contains 100 specially produced Japanese- and English-language songs~\cite{ref9}. Its AIST annotations provide beat structure, melody, and functional section intervals, with later contributions adding chords and other resources~\cite{ref10}. Such continued annotation efforts provide common repertoire for multiple tasks.

Isophonics provides chord, key, structural segmentation, and beat resources for popular and rock recordings by artists including the Beatles, Queen, and Michael Jackson, with annotation coverage varying across subsets~\cite{ref30}. Its Beatles subset contains 179 songs jointly annotated with chords, keys, sections, beats, and metrical positions.

Harmonix focuses on rhythm and structure analysis, providing manual beat, downbeat, and functional section annotations for 912 Western popular recordings, together with task benchmarks~\cite{ref5}. Professional musicians create annotations in a digital audio workstation, which are then converted to per-song JAMS and text files. Identifiers and onset information assist in locating and aligning recordings.

POP909 targets piano arrangement generation and aligns professional piano arrangements with the original audio of 909 popular songs~\cite{ref11}. Its main musical parts comprise vocal melody, lead melody, and piano accompaniment, alongside manual tempo curves and algorithmically extracted beats, downbeats, keys, and chords.

Annotation representations and interchange formats further support these resources. JAMS uses a unified JSON-based format to store annotations for different tasks and their metadata within one file~\cite{ref19}. Harte chord syntax provides a parsable representation of roots, chord qualities, and inversions~\cite{ref8}.

Table~\ref{tab:datasets} summarizes dataset size, temporal annotation coverage, and annotation sources.

\begin{table}[!t]
\centering
\caption{Dataset size and annotation coverage. $\checkmark$: detailed manual annotation or expert verification; $*$: annotation content or timing includes algorithmic generation or estimation; ---: not provided.}
\label{tab:datasets}
\footnotesize
\setlength{\tabcolsep}{2pt}
\begin{tabular*}{\columnwidth}{@{\extracolsep{\fill}}lrcccc@{}}
\toprule
Dataset & Songs & Phrase & Section & Chord & \shortstack{Beat/\\downbeat}\\
\midrule
SALAMI~\cite{ref3} & 1383 & $\checkmark$ & $\checkmark$ & --- & ---\\
Billboard 2.0~\cite{ref4} & 740 & $\checkmark$ & $\checkmark$ & $*$ & ---\\
\shortstack[l]{Isophonics--\\Beatles~\cite{ref30}} & 179 & --- & $\checkmark$ & $\checkmark$ & $\checkmark$\\
Harmonix~\cite{ref5} & 912 & --- & $\checkmark$ & --- & $\checkmark$\\
RWC-Pop~\cite{ref9,ref10} & 100 & --- & $\checkmark$ & $\checkmark$ & $\checkmark$\\
POP909~\cite{ref11} & 909 & --- & --- & $*$ & $*$\\
SongFormDB~\cite{ref7} & 14000+ & --- & $*$ & --- & ---\\
SongFormBench~\cite{ref7} & 300 & --- & $\checkmark$ & --- & ---\\
FaiSonic~MMD & 500 & $\checkmark$ & $\checkmark$ & $\checkmark$ & $\checkmark$\\
\bottomrule
\end{tabular*}
\end{table}

\subsection{Music Analysis Methods and Supervision}
Structure analysis typically evaluates boundary detection and functional labeling separately; descriptive perspective and temporal scale also affect label organization and evaluation~\cite{ref1,ref6}. SongFormer uses self-supervised features at multiple time scales and source embeddings to learn from structural data with heterogeneous coverage and label taxonomies~\cite{ref7}. Pretrained audio representations such as MuQ~\cite{ref18} and MusicFM~\cite{ref28} also provide reusable input features for downstream models.

Chord and rhythm models address harmonic context and periodic events, respectively. BTC uses a bidirectional Transformer to model chord sequences~\cite{ref16}, while ChordFormer combines Conformer modeling with structured chord representations for large-vocabulary recognition~\cite{ref24}. Beat and downbeat tracking approaches include neural networks combined with dynamic Bayesian networks~\cite{ref12} and state-space models for joint tempo and meter estimation~\cite{ref13}. BeatNet combines a convolutional recurrent network and particle filtering for online estimation~\cite{ref14}; Beat This! uses convolution, Transformers, and a shift-tolerant loss for beat and downbeat prediction~\cite{ref25}. The corresponding annotations provide training targets and evaluation references for these methods.

\looseness=-1
Joint modeling of temporal tasks also involves choices of time scale and supervision. All-In-One uses neighborhood attention to model metrical and functional structure and analyzes joint training through task ablations~\cite{ref15}; Beat-U organizes rhythm, chord, and structural predictions through downsampling, upsampling, and outputs at different resolutions~\cite{ref31}. SheetSage2 first learns local predictors from mixed supervision, uses structured decoding to produce complete event sequences, and then trains an autoregressive model with teacher-generated pseudo-labels~\cite{ref29}. These approaches explore joint use of temporal information through architecture, supervision acquisition, and decoding.

\section{Dataset Description}
\subsection{Annotation Procedure and Data Release}
\begin{figure*}[!t]
\centering
\includegraphics[width=\textwidth]{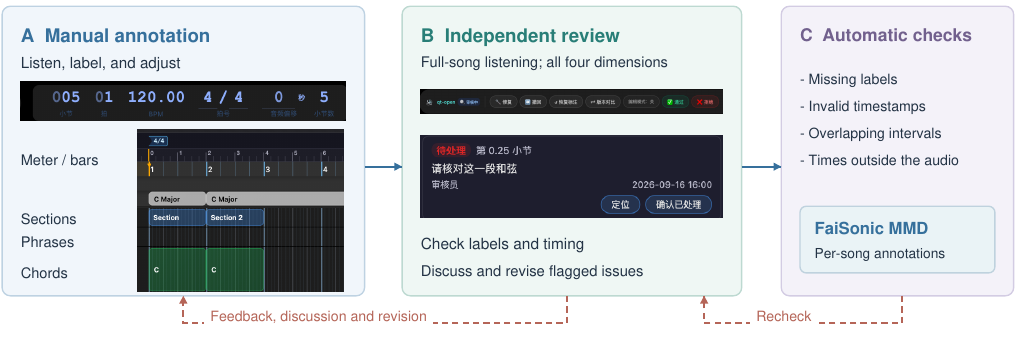}
\caption{Annotation and quality control workflow for FaiSonic~MMD. A: initial expert annotation; B: independent full-song review by a second expert; C: automatic checks. Solid arrows indicate forward flow; dashed arrows indicate revision or rechecking. Interface crops show anonymous demonstration data.}
\label{fig:annotation}
\end{figure*}

Figure~\ref{fig:annotation} summarizes the annotation and quality control workflow: initial expert annotation (A), independent review (B), and automatic checks (C). The annotator and reviewer hold at least a bachelor's degree from a music conservatory. Before beginning annotation, they receive shared training. This training follows unified annotation guidelines and accompanying documentation, establishing a common standard for annotation and review. Their roles remain fixed throughout: one produces the initial annotations, and the other checks them and provides feedback.

Both annotation and review use dedicated music annotation software integrating audio playback, waveform visualization, and multilayer label browsing. The annotator and reviewer can compare recordings, waveforms, and all four annotation layers within the same interface. Sine-wave chord playback and a metronome assist both experts in checking harmony and metrical positions.

Initial annotations are created song by song through listening. The same expert directly annotates phrases, sections, chords, and rhythm for each recording, completing all four dimensions within the same annotation process. No fixed order is prescribed across dimensions, and the annotator may revisit existing labels and adjust their content or temporal positions as work progresses.

The second expert, who does not participate in initial annotation, independently reviews all four dimensions for all 500 songs through full-song listening. Labels and temporal positions are checked against the recordings, with information across dimensions compared in the multilayer interface.

Problems identified during review are returned to the annotator. The two experts listen together, discuss the issues, and revise the annotations, corresponding to the revision path from review back to annotation in Figure~\ref{fig:annotation}.

Following human review, automatic checks detect missing labels, invalid timestamps, overlapping intervals, and times outside the audio duration. Detected anomalies are returned to the reviewer for further listening and inspection, corresponding to the path from automatic checks back to review in Figure~\ref{fig:annotation}. The resulting annotations are stored in per-song files; frame-level training labels are subsequently derived from the event and interval annotations.

We release FaiSonic~MMD annotations, metadata, and corresponding Mel spectrograms, together with code for feature extraction, training, inference, and evaluation. Mel spectrograms use extraction configurations compatible with pretrained BigVGAN~\cite{ref26} and DisCoder~\cite{ref27} models. Complete extraction parameters are provided, allowing researchers to reconstruct audio using these vocoders. Vocoder reconstruction is approximate, and the resulting waveforms are not guaranteed to match the original recordings exactly.

\subsection{Annotation Dimensions and Data Format}
FaiSonic~MMD contains 500 popular songs, with 250 songs each in Chinese and English, totaling approximately 32.86 hours. Each song has a JSON annotation file containing phrase intervals, section functions and intervals, chord roots and qualities with their intervals, and bar-start times with time signatures. Metadata include song title, performer, and duration.

All four dimensions use the start of the same recording version as time zero and share a common time unit. Intervals are represented by onset and duration, and events by their positions in the recording. The annotation layers can therefore be overlaid on a shared timeline to inspect which intervals contain particular events or to compare temporal offsets between boundaries. Each dimension retains its own granularity, and its temporal boundaries need not coincide with those of other dimensions.

\textbf{Phrases.} Phrase annotations record the start and end of local musical units, providing phrase intervals and boundary-detection targets. Phrases and sections are annotated separately. At a section transition, an anacrustic phrase may begin before the first bar of the following section.

\textbf{Sections.} Section annotations contain functional labels and their temporal intervals, covering intro, verse, pre-chorus, chorus, post-chorus, bridge, interlude, and outro. Each section instance is stored separately; adjacent instances may share the same functional label.

\textbf{Chords.} Chord roots and qualities are recorded separately. The 34 qualities cover triads, seventh chords, sixth chords, suspended chords, added-note chords, and extended chords including ninths, elevenths, and thirteenths. Annotation files retain the original qualities; training vocabularies and mappings are defined separately for the experimental models.

\textbf{Rhythm.} Rhythm annotations record manually marked bar-start times and their time signatures. Beats are obtained by uniformly subdividing the intervals between adjacent manually annotated bar starts according to the time signature. Bar starts also provide downbeat positions.

\subsection{Dataset Statistics}
\begin{figure}[!t]
\centering
\includegraphics[width=\columnwidth]{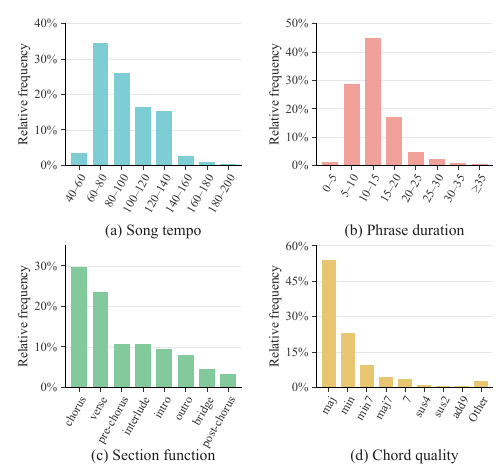}
\caption{FaiSonic~MMD statistics: (a) tempo in BPM, (b) phrase duration in seconds, (c) section function, and (d) chord quality. Percentages are normalized independently for each annotation dimension.}
\label{fig:statistics}
\end{figure}

Figure~\ref{fig:statistics} presents the distributions of tempo, phrase duration, section function, and chord quality in FaiSonic~MMD. All distributions are expressed as percentages. Tempo counts songs; the other distributions count phrases, section instances, and chord intervals, respectively.

\textbf{Tempo.} For 344 songs, the first enabled tempo control point supplies the representative BPM; the other 156 songs use the annotation's BPM field. Representative tempos range from 45.88 to 190.00 BPM, with a median of 88.00 BPM. Using left-closed, right-open intervals, the 60--80 and 80--100 BPM bins contain 173 and 130 songs, or 34.60\% and 26.00\%, respectively.

\textbf{Phrase duration.} The dataset contains 9,023 phrase intervals, with a median duration of 12.16 seconds and a range of 1.62--58.03 seconds. The most frequent bin, 10--15 seconds, contains 4,049 phrases (44.87\%); the 5--10 and 15--20 second bins contain 2,594 (28.75\%) and 1,543 (17.10\%), respectively. Together, these three bins cover 90.72\% of phrases. Only 28 phrases (0.31\%) last at least 35 seconds, so they are combined into a final $\geq35$-second bin.

\textbf{Section function.} Section annotations comprise 4,625 instances. Chorus and verse account for 29.64\% and 23.57\%, pre-chorus and interlude for 10.77\% and 10.72\%, intro and outro for 9.56\% and 8.04\%, and bridge and post-chorus for 4.48\% and 3.22\%, respectively. These proportions describe the number of occurrences of each function.

\textbf{Chord quality.} Chord annotations comprise 50,693 intervals, grouped by quality regardless of root or inversion. Major and minor account for 53.86\% and 23.25\%, while minor7, major7, and dominant7 account for 9.67\%, 4.33\%, and 3.47\%, respectively. The remaining individually displayed qualities, sus4, sus2, and add9, account for 1.16\%, 0.82\%, and 0.76\%. The 26 qualities individually below 0.5\% together account for 2.69\% and are grouped as Other in the figure. Labels maj, min, min7, maj7, 7, and add9 denote major, minor, minor seventh, major seventh, dominant seventh, and added-ninth major chords, respectively.

\section{Baseline Experiments}
The baseline experiments jointly supervise a comprehensive multitask model with all four annotation dimensions and evaluate six temporal prediction targets on the same data split, establishing reference performance for their joint use.

\subsection{Multitask Baseline Model}
\begin{figure*}[!t]
\centering
\includegraphics[width=.99\textwidth]{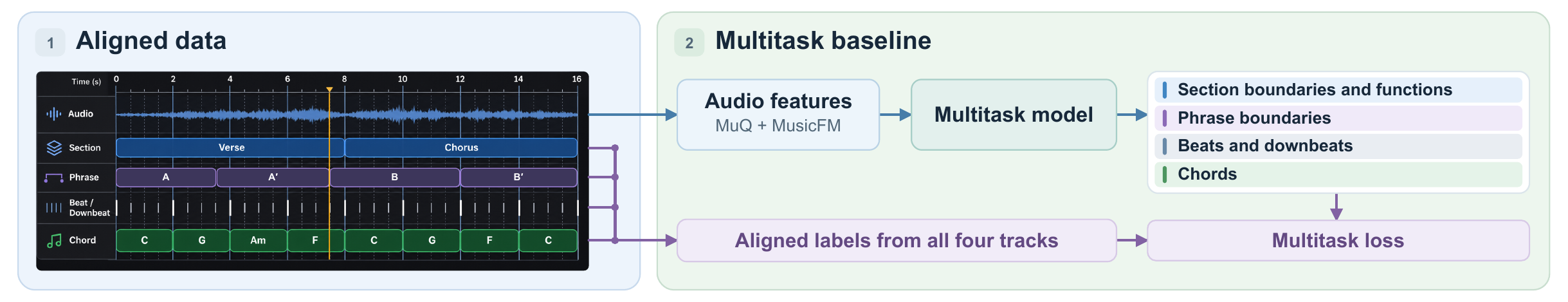}
\caption{Multitask training pipeline on FaiSonic MMD. (1) An illustrative view of the annotation editor shows section, phrase, beat and downbeat, and chord annotations on the recording timeline. (2) MuQ and MusicFM features feed the multitask model, and aligned labels from all four annotation dimensions supervise the corresponding predictions through the multitask loss.}
\label{fig:benchmark-pipeline}
\end{figure*}

Figure~\ref{fig:benchmark-pipeline} shows how each recording supplies audio features and aligned training targets. The targets share the recording timeline with the audio, so predictions can be compared with the corresponding annotations. The training objective combines losses across the supervised outputs.

The model uses frozen features from MuQ layer 6~\cite{ref18} and MusicFM layer 7~\cite{ref28}. Each recording is divided into consecutive, non-overlapping 30-second windows. Encoder outputs are concatenated along time within each stream to form full-song sequences, then concatenated across channels at corresponding time positions. Both streams have a frame rate of 25 Hz and 1,024 dimensions per frame. A linear layer maps the resulting 2,048-dimensional input to 256 dimensions, followed by two shared residual depthwise-separable convolutional blocks. Four independent task gates weight the shared representation at each frame and channel before passing it to the corresponding branch.

The phrase and rhythm branches each use a four-layer temporal convolutional network~\cite{ref20}. The chord branch applies two stride-2 convolutional layers to reach 6.25 Hz, followed by a two-layer Conformer variant~\cite{ref21} using frame-wise LayerNorm and global self-attention. The section branch applies three stride-2 layers to reach 3.125 Hz, followed by a four-layer Transformer encoder with dilated neighborhood attention (DiNA)~\cite{ref22}. Chord and section features are restored to 25 Hz by nearest-neighbor upsampling by factors of four and eight, respectively, implemented by repeating frames and cropping to the original sequence length. Both the section-function head and the boundary head, which compares multiscale contexts to the left and right of candidate positions, operate on the restored time grid.

The model uses independent prediction-head parameters for phrase boundaries, section boundaries, section functions, chords, beats, and downbeats. The rhythm branch has separate beat and downbeat outputs; their logits are added for beat prediction, while downbeat prediction retains its own output. The chord branch maps original qualities to 14 musical qualities plus no-chord N and out-of-vocabulary X, predicting 12 root classes and 16 quality labels separately. Appendix~\ref{app:chords} gives the mapping.

Beats and downbeats are obtained through confidence calibration, thresholding, and non-maximum suppression. Phrase and section boundaries use logistic regression to calibrate candidate scores from peak shapes and other tasks' predictions, followed by dynamic programming with soft duration constraints. Section-boundary decoding also penalizes changes in predicted section-function probabilities within a segment.

Section functions use Viterbi decoding~\cite{ref23}, with transition penalties adjusted by predicted section boundaries. Chord decoding has $1+12\times(14+1)=181$ states: one N state and the combinations of 12 roots with the 14 chord qualities and X. Joint Viterbi decoding uses predicted beats and downbeats as soft support for chord changes. All cross-task auxiliary information comes from model predictions.

\subsection{Training Configuration and Evaluation Protocol}
For our baseline experiments, FaiSonic~MMD is divided into 400 training, 50 validation, and 50 test songs. The training set contains 200 Chinese-language and 200 English-language songs; the validation and test sets each contain 25 Chinese-language and 25 English-language songs. The multitask baseline uses random seed 42 and epoch-12 weights; decoder parameters are fitted on validation data, frozen, and then applied to the test set.

Phrase and section boundaries are evaluated using F1 at tolerances of 0.5 and 3 seconds. Baseline section-function accuracy (ACC) is computed frame-wise over non-background reference frames pooled across test songs. Beat and downbeat F1 use one-to-one matching within 70 ms; baseline beat scoring excludes reference and predicted events in the first 5 seconds, whereas downbeat scoring uses the full recording.

Chord evaluation uses mir\_eval~\cite{ref17}, reporting Root, Majmin, Thirds, Triads, Sevenths, Tetrads, and MIREX. For the multitask baseline, reference qualities are mapped according to Appendix~\ref{app:chords}, then converted with predictions to evaluation symbols. Intervals with reference label X are excluded. Majmin and Sevenths also exclude unsupported reference qualities under their respective definitions. All metrics ignore inversions; scores are duration-weighted within each song and then averaged across songs.

\subsection{Results}
\begin{table}[!t]
\centering
\caption{Main multitask baseline results on FaiSonic~MMD and supplementary expanded-training results (\%). Differences are in percentage points.}
\label{tab:baseline}
\small
\setlength{\tabcolsep}{3pt}
\begin{tabular}{@{}lrrr@{}}
\toprule
Metric & \shortstack{FaiSonic\\MMD} & Expanded & Difference\\
\midrule
Phrase F1, 0.5 s & 62.90 & 63.32 & +0.42\\
Phrase F1, 3 s & 79.18 & 80.24 & +1.06\\
Section boundary F1, 0.5 s & 68.83 & 69.97 & +1.14\\
Section boundary F1, 3 s & 73.77 & 75.57 & +1.80\\
Section function ACC & 73.33 & 76.44 & +3.11\\
Beat F1 & 86.18 & 87.97 & +1.79\\
Downbeat F1 & 81.40 & 84.48 & +3.08\\
Chord Root & 80.16 & 83.52 & +3.36\\
Chord Majmin & 75.71 & 80.94 & +5.22\\
Chord Thirds & 75.72 & 80.45 & +4.72\\
Chord Triads & 73.94 & 78.44 & +4.50\\
Chord Sevenths & 65.66 & 69.79 & +4.12\\
Chord Tetrads & 64.27 & 68.32 & +4.05\\
Chord MIREX & 74.93 & 79.03 & +4.10\\
\bottomrule
\end{tabular}
\end{table}

Table~\ref{tab:baseline} reports the multitask baseline's main results for the six prediction targets on FaiSonic~MMD.

Joint training yields phrase-boundary F1 of 62.90\% and 79.18\% at 0.5- and 3-second tolerances, section-boundary F1 of 68.83\% and 73.77\%, and section-function ACC of 73.33\%. Beat and downbeat F1 reach 86.18\% and 81.40\%, while chord Root, Majmin, and MIREX are 80.16\%, 75.71\%, and 74.93\%, respectively. The baseline performs all six temporal prediction tasks on the fixed split of the 500-song dataset, establishing reference performance for joint supervision with all four annotation dimensions.

The supplementary experiment adds expert-annotated recordings from our larger existing collection to expand training to 2,350 songs, keeping validation and test sets unchanged. All reported metrics improve, showing that this baseline benefits from additional expert-annotated training material of the same kind.

\section{Conclusion and Discussion}
We present FaiSonic~MMD for temporal music analysis, jointly providing expert phrase, section, chord, and rhythm annotations for 500 popular recordings. Song-by-song annotation and independent review verify musical content and temporal positions, while a shared recording-based time reference allows direct comparison across annotation granularities. The public dataset and comprehensive multitask baseline provide multidimensional annotations and a reference implementation for their joint use, with baseline results establishing a reusable evaluation starting point for subsequent methods.

Multidimensional annotations can support the analysis of musical relationships and sources of error beyond prediction tasks. For example, researchers can compare phrase durations across section functions, examine the positions of chord changes within bars, or analyze temporal offsets between phrase and section boundaries. These relationships can motivate hypotheses about context and event positions, which can then be tested through modeling on fixed splits. Annotation dimensions can also be combined or selected individually according to the research objective.

The comprehensive baseline provides one approach to using all four annotation dimensions jointly. Its targets involve local events, sustained intervals, and larger-scale structural organization. Future work can use the same annotations to compare choices of temporal resolution, context range, and decoding strategy. Grouped error analyses using existing tempo and section-function annotations can further examine model performance in different musical contexts and inform task design and repertoire expansion.

All metrics improve in the supplementary expanded-training experiment, providing empirical support for further collection of expert-annotated material of the same kind. Keeping the evaluation set fixed would allow future work to examine how repertoire size and content coverage affect different prediction targets.

The dataset remains limited in scale and coverage, focusing on Chinese- and English-language popular recordings. Its applicability to other languages, musical styles, and recording conditions requires further evaluation. The baseline also provides a starting point for broader method comparisons and cross-dataset research.

\section{Acknowledgments}
Hangzhou FaiSonic Intelligent Music Technology Co., Ltd.\ funded the annotation of FaiSonic~MMD and provided the annotation software. We thank the professional musicians who contributed to annotation and review.

\FloatBarrier

\appendix
\section{Chord-Quality Mapping for the Multitask Baseline}
\label{app:chords}
The baseline maps original chord qualities to 14 training qualities, plus no-chord N and out-of-vocabulary X. Roots remain 12 pitch classes, indexed from C as 0 through B as 11.

\subsection{Quality Mapping}
The table lists the normalized input qualities for each training class, with the training-class ID in parentheses.
\begin{center}
\setlength{\tabcolsep}{3pt}
\begin{tabular}{@{}lp{.64\columnwidth}@{}}
\toprule
Training class (ID) & Normalized input qualities\\
\midrule
N (0) & N\\
major (1) & major, add9, add11, add13\\
minor (2) & minor, minor-add9, minor-add11, minor-add13\\
diminished (3) & diminished\\
augmented (4) & augmented\\
major7 (5) & major7, major9, major11, major13\\
minor7 (6) & minor7, minor9, minor11, minor13\\
dominant7 (7) & dominant7, dominant9, 11, 13, altered\\
half-dim7 (8) & half-dim7\\
dim7 (9) & dim7\\
major6 (10) & major6, major69\\
minor6 (11) & minor6\\
sus2 (12) & sus2\\
sus4 (13) & sus4\\
minor-major7 (14) & minor-major7\\
X (15) & X, power, dominant7sus4, dominant7sus2, dominant9sus4\\
\bottomrule
\end{tabular}
\end{center}
\end{document}